\documentclass[a4paper,12pt]{article}
\usepackage[dvips]{graphicx}
\usepackage{multirow}
\usepackage{longtable}
\usepackage[usenames]{color}
\usepackage{threeparttable}
\usepackage[english]{babel}
\usepackage[aboveskip=5pt,margin=5pt,belowskip=5pt]{caption}
\usepackage{geometry} 
\usepackage{hyperref}
\begin{document}

\title{Low-Density Structures in the Local Universe. \\
II. Nearby Cosmic Voids}

\author{A. Elyiv$^{1,2}$, I. Karachentsev$^3$, V. Karachentseva$^1$, O. Melnyk$^{2,4}$, D. Makarov$^3$}

\date{}

\maketitle

{\footnotesize \it \noindent 

$^1$ Main Astronomical Observatory, National Academy of Sciences, Kiev, 03680 Ukraine \\
$^2$ Institut d'Astrophysique et de Geophysique, Universit\'{e} de Li\`{e}ge, Li\`{e}ge, B5C B4000 Belgium \\
$^3$ Special Astrophysical Observatory of the Russian AS, Nizhnij Arkhyz 369167, Russia \\
$^4$ Astronomical Observatory, Taras Shevchenko National University of Kiev, 04053 Ukraine \\
}

\begin{abstract}
We present the results of the search for spherical volumes
containing no galaxies with luminosities brighter than the
Magellanic Clouds in the Local Supercluster and its vicinity.
Within a distance of 40 Mpc from us, 89 cosmic voids were
discovered with the diameters of 24 to 12~Mpc, containing no
galaxies with absolute magnitudes brighter than $M_K < -18.4$. A
list of these voids and the sky distribution maps are given. It
was found that 93\% of spherical voids overlap, forming three more
extended percolated voids (hypervoids). The largest of them, HV1,
has 56 initial spherical cells and extends in a horseshoe shape,
enveloping the Local Volume and the Virgo cluster. The Local Void
(Tully, 1988) in the \mbox{Hercules--Aquila} region is the closest
part of the HV1. Another hypervoid, HV2, contains 22 spherical
voids in the Eridanus constellation, and the third compact
hypervoid (HV3) comprises 6 spherical cells in the Bootes. The
total volume of these voids incorporates about 30\% of the Local
Universe. Among 2906 dwarf galaxies excluded from the original
sample ($n = 10502$) in the search for spherical volumes, only 68
are located in the voids we have discovered. They are
characterized by late morphological types (85\% are Ir, Im, BCD,
Sm), absolute magnitudes $M_B$ ranging from $-13.0$ to $-16.7$,
moderate star formation rates ($\log {\rm SSFR}\sim
-10~M_{\odot}\,{\rm yr}^{-1} L_{\odot}^{-1}$) and gas reserves per
luminosity unit twice to three times larger than in the other
dwarf galaxies located in normal environments. The dwarf
population of the voids shows a certain tendency to sit shallow
near the surfaces of cosmic voids.\\
\textbf{Keywords:} cosmology: large-scale structure of Universe.
\end{abstract}

\maketitle

\section{INTRODUCTION}

The main element in the picture of the large-scale structure of
the Universe, often referred to as the ``Cosmic Web'',  are the
vast low-density regions, separated by the ``walls'' and
filaments. The first observational evidence for the existence of
giant cavities (voids) appeared about 30 years
ago~\cite{Joe1978:Elyiv_n,Greg1978:Elyiv_n,Kir1981:Elyiv_n}, but
the concept of cosmic voids was implemented in common practice
with the advent of mass surveys of galaxy redshifts. According to
the current estimates, the sizes of voids span a wide range of
extents from the supervoids with the diameters of about
200~Mpc~\cite{Lin1995:Elyiv_n} to mini-voids covering about
3--5~Mpc~\cite{Tikh2006:Elyiv_n}. There is a conception that the
typical number density of galaxies in cosmic voids is at least an
order of magnitude lower than global mean density.

A lot of studies can be found in the literature, examining the
statistics of sizes and shapes of voids, as well as the features
of their population. A recent review of these studies was
presented by van~de~Weygaert and Platen~\cite{van2009:Elyiv_n}.
Different authors have used various algorithms for identifying the
voids in mass sky surveys. Some of them suggest a complete absence
of galaxies in the voids up to a fixed luminosity level. Other
criteria allow the possibility of presence of a small number of
galaxies with normal luminosity in the low-density regions. In the
latter case, these regions should rather be called lacunas instead
of voids.

Two important questions still remain unresolved: 1) whether there
are volumes of space, completely devoid of galaxies, 2) if there
are any signs of the void expansion in the kinematics of galaxies
surrounding it. The answers to these questions are closely related
to the choice of the most plausible scenario of the formation of
the large-scale structure of the
Universe~\mbox{\cite{Peeb2001:Elyiv_n,van2009:Elyiv_n}}. It is
evident that the best opportunity to explore the dwarf population
of the voids and the kinematics of galaxies around them is
provided by the nearest voids. Compiling the atlas of nearby
galaxies, Tully and Fisher~\cite{Tul1987:Elyiv_n} have discovered
a giant empty region in the \mbox{Aquila--Hercules}
constellations, which starts immediately at the threshold of our
Local Group of galaxies, and occupies about a quarter of the
entire sky. Inside this Local Void only two dwarf galaxies were
found to date: KK~246~\cite{Kar2006:Elyiv_n} and
\mbox{ALFAZOA~J1952+1428~\cite{McIn2011:Elyiv_n}} with absolute $B$-magnitudes of $-13.7^{m}$ and $-13.5^{m}$,
respectively. The analysis of the data on the radial velocities
and distances of the galaxies in the vicinity of the Local Void
points to an expansion of its frontiers at a velocity of about
300~km/s~\cite{Nas2011:Elyiv_n}.

Using the 2dFGRS \cite{Col2011:Elyiv_n} and SDSS\footnote{Sloan
Digital Sky Survey
\tt{(http://www.sdss.org)}.}~\cite{Abaz2009:Elyiv_n} galaxy
redshift surveys, Patiri et al.~\cite{Pat2006:Elyiv_n} and Hoyle
et al.~\cite{Hoy2012:Elyiv_n} have identified a large number of
distant voids at the characteristic distances of $z\sim0.1$.
However, we have not found in the literature any systematic lists
of voids in a closer volume on the scale of $z\sim0.01$. An
enumeration of nearby voids in the southern and northern sky,
indicating their rough outlines can be found in the Fairall's
manuscript~\cite{Fai1988:Elyiv_n}.  Pustilnik and
Tepliakova~\cite{Pus2011:Elyiv_n} have investigated the properties
of dwarf galaxies in the region of the Lynx--Cancer void.
Saintonge et al.~\cite{Sain2008:Elyiv_n} noted the presence of a
nearby $(V\sim2000$ km/s) void in the Pisces constellation
according to the ALFALFA survey data~\cite{Gio2005:Elyiv_n}.

\begin{figure}[]
\includegraphics[width=0.90\columnwidth]{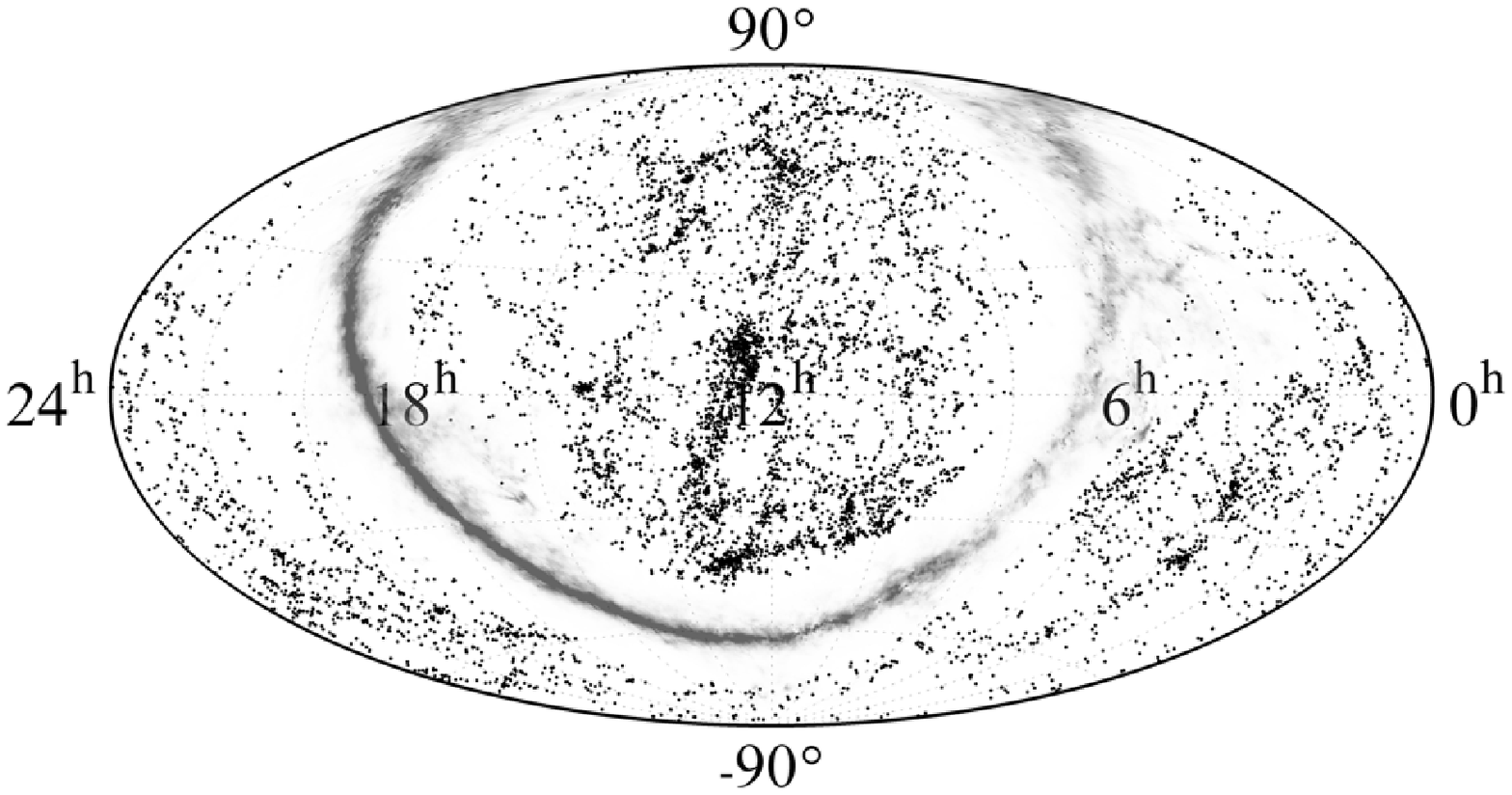}\\
\includegraphics[width=0.90\columnwidth]{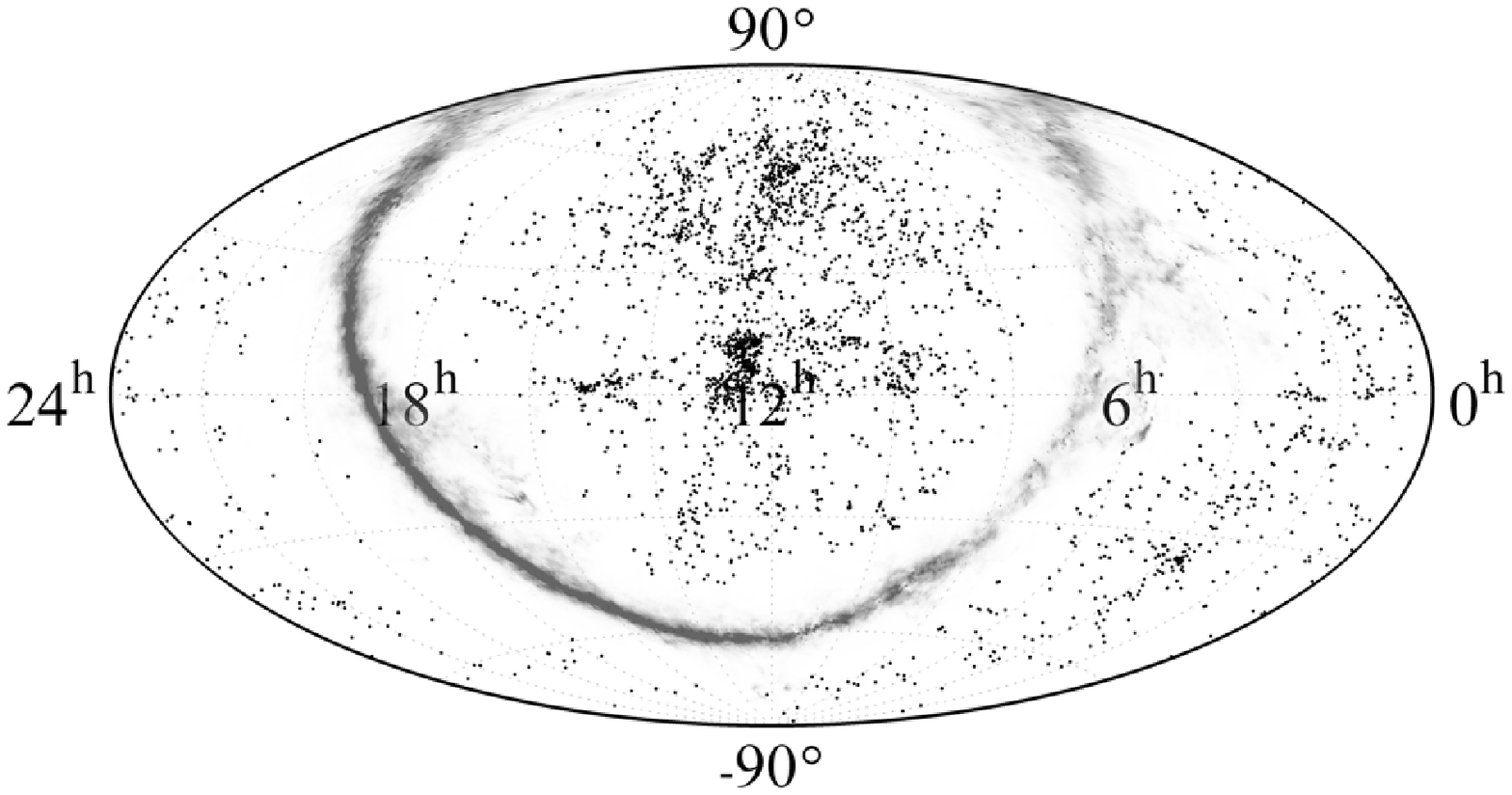}
\caption{The distribution of 7596 bright
(top panel) and 2096 dwarf (bottom panel) galaxies with
\mbox{$V_{\rm LG} =0$--$3500$~km/s} in equatorial coordinates. The
region of significant Galactic absorption with $A_B>2.0^{m}$ is described by a ragged gray stripe.}
\end{figure}

In the first paper of this series~\cite{Karach2012:Elyiv_n},
applying the percolation method we have identified the diffuse
agglomerates of galaxies in the Local Supercluster, which are
located in the regions of low matter density, and discussed the
properties of the most populated non-virialized structures.
In this paper we describe the algorithm for identifying the voids in the Local
Supercluster, give a list of nearby (within 40 Mpc) voids and
briefly discuss the properties of dwarf galaxies located in the
nearby voids.

\section{ORIGINAL SAMPLE OF GALAXIES}

To search for voids in the Local Universe we used a sample of
10\,502 galaxies with radial velocities  \mbox{$0<V_{\rm
LG}<3500$~km/s} relative to the centroid of the Local Group. The
sample includes both the northern and southern sky, except for the
low galactic latitudes  $|\,b\,|<15^{\circ}$. For these galaxies,
presented in recent versions of the HyperLeda\footnote{Lyon
Database for Physics of
Galaxies\\
\tt{(http://leda.univ-lyon1.fr)}.}~\cite{Pat2003:Elyiv_n} and
NED\footnote{NASA/IPAC Extragalactic Database\\
\tt{(http://nedwww.ipac.caltech.edu)}.} databases we have refined
their angular dimensions and morphological types. Since the stellar
mass of the galaxy is best expressed through its \mbox{$K$-band}
luminosity, we have used the \mbox{$K_s$-luminosities} of galaxies
from the 2MASS survey~\cite{Jar2000:Elyiv_n}. In the absence of
these data, we determined the apparent \mbox{$K$-magnitudes} from
the apparent \mbox{$B$-magnitudes} and average color index
\mbox{$\langle B-K \rangle$} separately for each morphological
type according to~\cite{Jar2000:Elyiv_n}. From the original
sample, we have excluded faint galaxies with $K>15.0^{m}$, and
objects with negative radial velocities. At the  Hubble parameter
of $H_0=73$ km/s/Mpc the distance modulus for the far boundary of
our sample corresponds to $m-M=33.4^{m}$. In order to have the same
conditions for the void identification  both in the near and far
regions of our volume, we have excluded from the analysis the
dwarf galaxies with absolute magnitudes fainter than $M_K=-18.4$.
This threshold is roughly equivalent to the luminosity of the
Small Magellanic Cloud-type dwarfs. At that, the distances to the
galaxies were determined by their radial velocity with the $H_0$
given above.

The top and bottom panels of Fig.~1 show the sky distribution of
7596 bright galaxies with \mbox{$M_K<-18.4$} and 2906 dwarf
galaxies, excluded from the void identification procedure,
respectively. Both sub-samples show the effects of clustering into
groups and clusters, as well as the concentration to the equator
of the Local Supercluster. The distribution of dwarfs is affected
by their over-clustering in the volumes of the  nearby Virgo
cluster and the Canes~Venatici~I cloud, as well as  high density
of galaxies with known redshifts in the SDSS
survey~\cite{Abaz2009:Elyiv_n} region.

\section{VOID IDENTIFICATION ALGORITHM}

For each of the 7596 galaxies with $M_K<-18.4$ within $D=V_{\rm
LG}/73=48$ Mpc the Cartesian equatorial coordinates X, Y, Z were
calculated. Then, in this volume, applying the galactic latitude
restriction, we searched for the maximum radius $R$ for a sphere
which does not contain any galaxies. For this purpose, we have
carried out a search of all possible coordinates of the
circumcenter and the radius of the sphere. The search increment
has been set at 1.5~Mpc for the elapsed time reasoning (a
compromise between the accuracy and required computer time). The
limiting conditions were set so that the center of the sought void
would lie within $D=48$~Mpc and not be located in the cone of the
Milky Way $|\,b\,|<15^{\circ}$.

Further, we looked for the following void with the maximum radius
and containing no galaxies. The limiting conditions were
supplemented with a new constraint: the center of the desired void
has to be located outside the volume, occupied by the previous
void. This procedure was repeatedly iterated accounting for the
locations and sizes of all previous voids. This process was in
progress until the number of voids has reached \mbox{$n\geq 100$.}
The result is a collection of 179~empty spherical volumes with the
radii of 12 to 6~Mpc, quite a few of which partially overlap each
other.

The applied algorithm does obviously include several parameters,
the choice of which affects the final list of voids. One of them
is the minimum distance of the void center to the boundaries of
the volume, as well as the minimum distance between the centers of
the voids. Another parameter is the  threshold absolute magnitude
of the dwarf galaxies \mbox{($M_K=-18.4$),} the possible presence
of which in the empty volume is ignored. The third parameter is
the minimum radius of the spherical void~\mbox{($R_{\rm min}=6.0$
Mpc),} which terminates the application of the algorithm.

Attempting to reject the sphericity of the identified empty
volumes, as did Tikhonov and Karachentsev~\cite{Tikh2006:Elyiv_n}
significantly complicates the algorithm. In addition, a comparison
of the SDSS~DR7 data with the galaxy distribution of the
Millennium~1 model catalog  by Tavasoli et
al.~\cite{Tav2012:Elyiv_n} has shown that the shape of voids tends
to be spherical. If needed,   nonspherical voids can be obtained
within our approach by joining two or more intersecting spherical
volumes in their association: a ``dumbbell'', a ``boomerang'' or a
`` string''.

\section{A LIST OF EMPTY VOLUMES IN \\ THE LOCAL UNIVERSE}

The results of our search for nearby spherical voids applying the
above algorithm are presented in Table~1. The table columns
include: (1) the number of the void in the procedure we have
adopted, when each subsequent step yields the voids of an ever
smaller radius; (2) the distance to the center of the void in Mpc; (3, 4) equatorial coordinates of the void's center in
degrees; (5, 6) the linear and  angular diameter of the void; (7)
notes, which stipulate the location of 12 most nearby voids in the
constellations, and the membership of the given empty volume in
the ever more extended formations, called the hypervoids (HV).

\begin{table*}
\setcaptionmargin{0mm} 
\caption{The list of spherical voids in the Local Universe}

{\scriptsize
\begin{tabular}{r|c|r|r|r|c|cl||r|c|r|r|r|c|cl}

\hline
\multicolumn{1}{c|}{No.} & $D$,& \multicolumn{1}{c|}{RA,}& \multicolumn{1}{c|}{Dec,}  &  \multicolumn{1}{c|}{$R$,}&$r$, &\multicolumn{2}{c||}{Note} & \multicolumn{1}{c|}{No.} & $D$,& \multicolumn{1}{c|}{RA,}& \multicolumn{1}{c|}{Dec,}  &  \multicolumn{1}{c|}{$R$,}&$r$, &\multicolumn{2}{c}{Note}\\
   & Mpc &\multicolumn{1}{c|}{deg}& \multicolumn{1}{c|}{deg} & \multicolumn{1}{c|}{Mpc}  & deg &   &  &       & Mpc &\multicolumn{1}{c|}{deg}& \multicolumn{1}{c|}{deg} & \multicolumn{1}{c|}{Mpc}  & deg &  &       \\
\hline
\multicolumn{1}{c|}{(1)}& (2)& \multicolumn{1}{c|}{(3)}& \multicolumn{1}{c|}{(4)}& \multicolumn{1}{c|}{(5)}& (6)&\multicolumn{2}{c||}{(7)} & \multicolumn{1}{c|}{(1)}& (2)& \multicolumn{1}{c|}{(3)}& \multicolumn{1}{c|}{(4)}& \multicolumn{1}{c|}{(5)}& (6)&\multicolumn{2}{c}{(7)}\\
\hline

38    & 8.7 & 301 &   0   & 7.5  &  59 & HV1,&  Aqr          &  47  &30.9 &   8 &  47   & 7.5  &  14 & \multicolumn{2}{c}{HV1}                \\
 9    &16.1 & 288 & --28  & 9.0  &  34 & HV1,& Sgr           & 122  &31.2 & 279 &  29   & 6.0  &  11 & \multicolumn{2}{c}{HV1}                \\
 116  &16.6 &  85 &  --5  & 6.0  &  21 & HV1,& Ori           &  26  &31.5 & 141 & --72  & 7.5  &  14 & \multicolumn{2}{c}{HV2}                \\
  34  &17.6 & 275 &  20   & 7.5  &  25 & HV1,& Her           & 161  &31.8 &  16 & --28  & 6.0  &  11 & \multicolumn{2}{c}{HV2}                \\
  27  &18.4 & 117 &  24   & 7.5  &  24 & HV1,& Gem--Leo      & 117  &31.9 & 278 & --45  & 6.0  &  11 & \multicolumn{2}{c}{HV1}                \\
 136  &19.1 & 360 &  45   & 6.0  &  18 & HV1,& And           & 159  &32.0 & 329 &  --8  & 6.0  &  11 & \multicolumn{2}{c}{HV1}                \\
 147  &20.1 & 347 &   4   & 6.0  &  17 &     & Psc--Peg      & 100  &32.1 & 258 &  25   & 6.0  &  11 & \multicolumn{2}{c}{HV1}                \\
 144  &21.0 &  18 &  25   & 6.0  &  17 & HV1,& Psc           & 114  &32.3 &  83 &  68   & 6.0  &  11 & \multicolumn{2}{c}{HV1}                \\
  90  &21.5 & 238 & --25  & 6.0  &  16 & HV1,& Sco--Lib      &  35  &32.4 &  79 & --76  & 7.5  &  13 & \multicolumn{2}{c}{HV2}                \\
 119  &21.5 & 283 &  50   & 6.0  &  16 & HV1,& Dra           & 146  &32.4 & 347 & --52  & 6.0  &  11 & \multicolumn{2}{c}{HV2}                \\
 135  &21.8 & 309 &  12   & 6.0  &  16 & HV1,& Del           & 148  &32.7 &  35 &  43   & 6.0  &  11 & \multicolumn{2}{c}{HV1}                \\
  31  &23.9 & 248 & --70  & 7.5  &  18 & HV1,& Aps           & 157  &32.8 &  32 & --30  & 6.0  &  11 & \multicolumn{2}{c}{HV2}                \\
 120  &25.3 &  68 & --71  & 6.0  &  14 & \multicolumn{2}{c||}{HV2}                & 156  &32.9 &  29 &  33   & 6.0  &  11 & \multicolumn{2}{c}{HV1}                \\
 107  &25.6 & 265 &  50   & 6.0  &  14 & \multicolumn{2}{c||}{HV1}                &   2  &33.1 & 339 &  39   &12.0  &  21 & \multicolumn{2}{c}{HV1}                \\
 118  &25.6 & 279 & --45  & 6.0  &  14 & \multicolumn{2}{c||}{HV1}                &  49  &33.3 & 319 &  27   & 7.5  &  13 & \multicolumn{2}{c}{HV2}                \\
 130  &26.0 & 299 & --44  & 6.0  &  13 & \multicolumn{2}{c||}{HV1}                & 165  &33.5 &  22 & --16  & 6.0  &  10 & \multicolumn{2}{c}{HV2}                \\
  39  &26.1 &  80 &   7   & 7.5  &  17 & \multicolumn{2}{c||}{HV1}                &  56  &33.7 &   3 & --21  & 7.5  &  13 & \multicolumn{2}{c}{HV1}                \\
 123  &26.6 & 288 &  58   & 6.0  &  13 & \multicolumn{2}{c||}{HV1}                & 124  &34.0 & 291 & --68  & 6.0  &  10 & \multicolumn{2}{c}{HV1}                \\
  42  &27.0 &  60 &  26   & 7.5  &  16 & \multicolumn{2}{c||}{HV1}                & 166  &34.3 & 337 &  --3  & 6.0  &  10 & \multicolumn{2}{c}{HV1}                \\
 140  &27.1 &  29 &  46   & 6.0  &  13 & \multicolumn{2}{c||}{HV1}                &  51  &34.6 &   3 & --34  & 7.5  &  13 & \multicolumn{2}{c}{HV2}                \\
 111  &27.6 &  90 & --45  & 6.0  &  13 & \multicolumn{2}{c||}{HV2}                &  13  &34.7 &  50 &  18   & 9.0  &  15 & \multicolumn{2}{c}{HV1}                \\
  96  &27.7 & 249 &  22   & 6.0  &  13 & \multicolumn{2}{c||}{HV1}                &  40  &34.7 &  77 &  12   & 7.5  &  13 & \multicolumn{2}{c}{HV1}                \\
 154  &27.7 &  22 &  29   & 6.0  &  12 & \multicolumn{2}{c||}{HV1}                &   3  &34.7 &  59 & --60  &10.5  &  18 & \multicolumn{2}{c}{HV2}                \\
 143  &28.0 &  14 & --49  & 6.0  &  12 & \multicolumn{2}{c||}{HV2}                & 104  &34.8 & 263 &   0   & 6.0  &  10 & \multicolumn{2}{c}{HV1}                \\
 150  &28.0 & 319 &  --9  & 6.0  &  12 & \multicolumn{2}{c||}{HV1}                & 112  &35.0 &  90 & --47  & 6.0  &  10 & \multicolumn{2}{c}{HV2}                \\
  99  &29.0 & 254 & --15  & 6.0  &  12 & \multicolumn{2}{c||}{HV1}                &  44  &35.4 &  65 &  10   & 7.5  &  12 & \multicolumn{2}{c}{HV1}                \\
 151  &29.1 &  23 & --38  & 6.0  &  12 & \multicolumn{2}{c||}{HV2}                &  75  &35.4 & 205 &  10   & 6.0  &  10 & \multicolumn{2}{c}{HV3}                \\
  86  &29.5 & 125 &  --9  & 6.0  &  12 &     &               &  85  &35.5 & 239 & --10  & 6.0  &  10 & \multicolumn{2}{c}{HV1}                \\
  79  &29.7 & 204 &   6   & 6.0  &  12 & \multicolumn{2}{c||}{HV3}                & 108  &35.6 & 265 &  62   & 6.0  &  10 & \multicolumn{2}{c}{HV1}                \\
   8  &30.1 & 101 &  40   & 9.0  &  17 & \multicolumn{2}{c||}{HV1}                & 141  &35.6 &  60 & --22  & 6.0  &  10 & \multicolumn{2}{c}{HV2}                \\
  95  &30.5 & 252 &  11   & 6.0  &  11 & \multicolumn{2}{c||}{HV1}                &  72  &35.7 & 158 &   0   & 6.0  &  10 & \phantom{HV1,} & \phantom{Gem--Leo }   \\

\hline
\end{tabular}}
\end{table*}

\setcounter{table}{0}
\begin{table}
\caption{(Contd.)}
\medskip
{\scriptsize
\begin{tabular}{r|c|r|r|r|r|c}
\hline
\multicolumn{1}{c|}{No.} & $D$,& \multicolumn{1}{c|}{RA,}& \multicolumn{1}{c|}{Dec,}  &  \multicolumn{1}{c|}{$R$,}&\multicolumn{1}{c|}{$r$,} &Note\\
   & Mpc &\multicolumn{1}{c|}{deg}& \multicolumn{1}{c|}{deg} & \multicolumn{1}{c|}{Mpc}  & \multicolumn{1}{c|}{deg} &      \\
\hline
\multicolumn{1}{c|}{(1)}& (2)& (3)& (4)& (5)& (6)&(7)\\
\hline
 106  &35.8 & 257 & --67  & 6.0  &  10 & HV1                \\
  15  &35.9 & 324 &  10   & 9.0  &  15 & HV1                \\
 102  &35.9 & 106 &  53   & 6.0  &  10 & HV1                \\
 145  &35.9 & 303 &  --5  & 6.0  &  10 & HV1                \\
   5  &36.2 & 306 &  12   &10.5  &  17 & HV1                \\
 139  &36.7 & 294 & --10  & 6.0  &   9 & HV1                \\
   4  &37.2 &  40 & --47  &10.5  &  17 & HV2                \\
 105  &37.4 & 263 &  50   & 6.0  &   9 & HV1                \\
  80  &37.6 & 225 &  16   & 6.0  &   9 & HV3                \\
  18  &37.6 & 152 &  31   & 7.5  &  11 & \phantom{HV1, Gem--Leo} \\
 109  &37.7 & 117 & --85  & 6.0  &   9 & HV2                \\
  11  &37.7 & 288 & --23  & 9.0  &  14 & HV1                \\
  12  &37.9 &  56 &  31   & 9.0  &  14 & HV1                \\
  91  &38.0 & 210 &  71   & 6.0  &   9 &                    \\
  53  &38.1 & 329 &  23   & 7.5  &  11 & HV1                \\
 126  &38.1 &  27 & --80  & 6.0  &   9 & HV2                \\
  60  &38.3 &   2 & --11  & 7.5  &  11 & HV2                \\
 168  &38.3 &   0 &  31   & 6.0  &   9 & HV1                \\
 173  &38.4 &   5 &  11   & 6.0  &   9 &                    \\
  97  &38.6 & 257 &  33   & 6.0  &   9 & HV1                \\
 160  &38.7 &  13 & --44  & 6.0  &   9 & HV2                \\
 115  &39.0 &  84 &  67   & 6.0  &   9 & HV1                \\
  48  &39.1 &  51 & --32  & 7.5  &  11 & HV2                \\
  74  &39.4 & 216 &  11   & 6.0  &   9 & HV3                \\
 153  &39.4 &  18 & --53  & 6.0  &   9 & HV2                \\
  54  &38.5 &  39 &  13   & 7.5  &  11 & HV1                \\
  69  &39.7 & 209 &   4   & 6.0  &   9 & HV3                \\
\hline
\end{tabular}}
\end{table}

It should be noted that the table shows only 89~voids, ranked
according to the distance from the observer up to 40 Mpc, from the
total number of 179.\footnote{The whole list of 179 voids can be
obtained upon request to the first author.} We have excluded from
the list a half of the most distant voids based on the following
reasoning. The distribution of the integral number of voids
depending on the distance of their centers $D$ shows that near the
far border of the considered volume in the range of
\mbox{$D=40$--$48$~Mpc,} there is approximately a double excess of
voids, as compared with the homogeneous distribution \mbox{$n\sim
D^3$.} This excess is due to the decrease of the number density of
galaxies with measured radial velocities in the most distant parts
of the Local Universe. In addition, the lack of data on the
galaxies outside the radius of~48 Mpc increases the probability of
finding an empty volume and ``adhesion'' of these excessive voids
on the far boundary.


\begin{figure}[]
\vspace{7mm}
\setcaptionmargin{3mm} \hspace{-1.5mm} 
\includegraphics[scale=0.3]{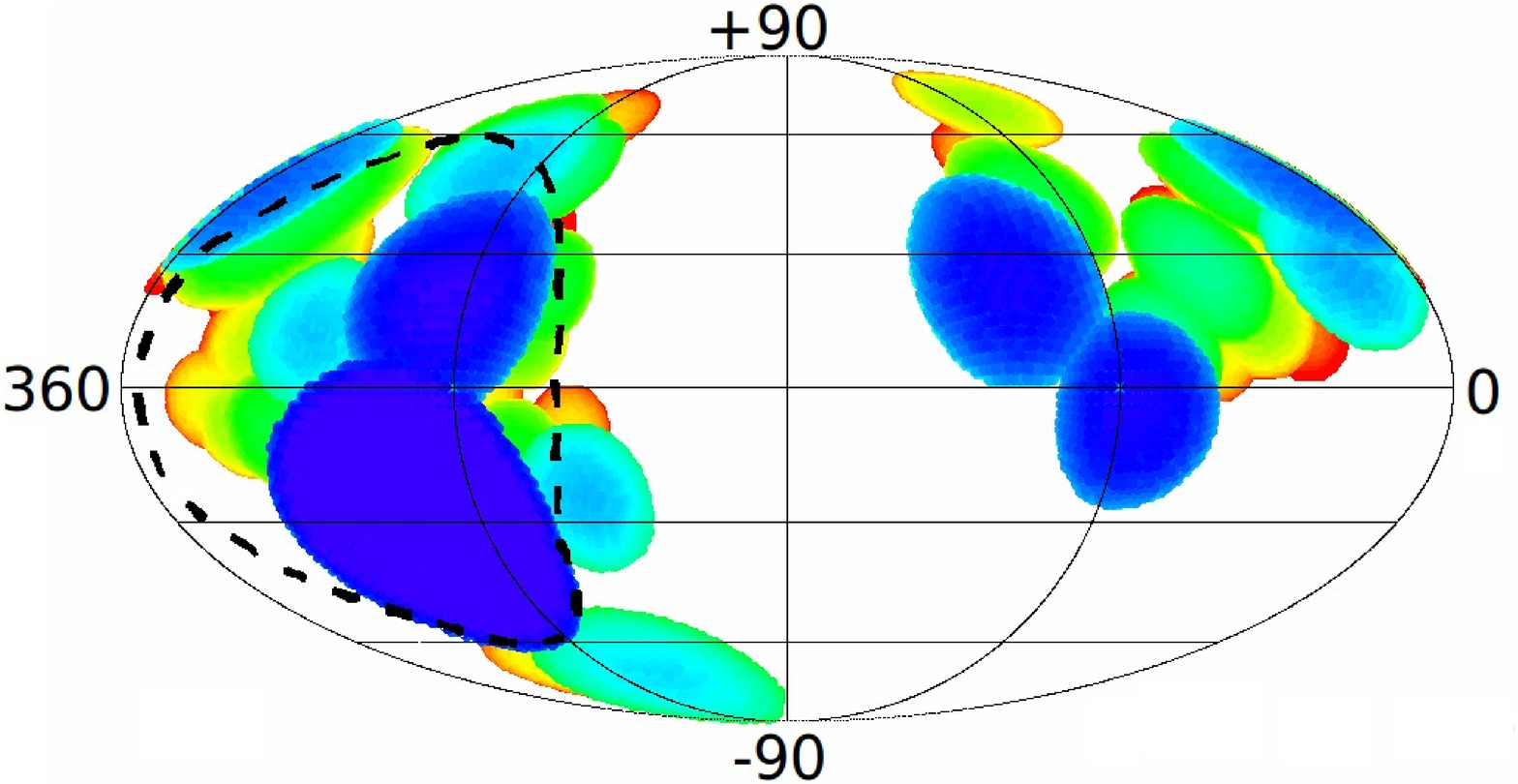}\\
\includegraphics[scale=0.3]{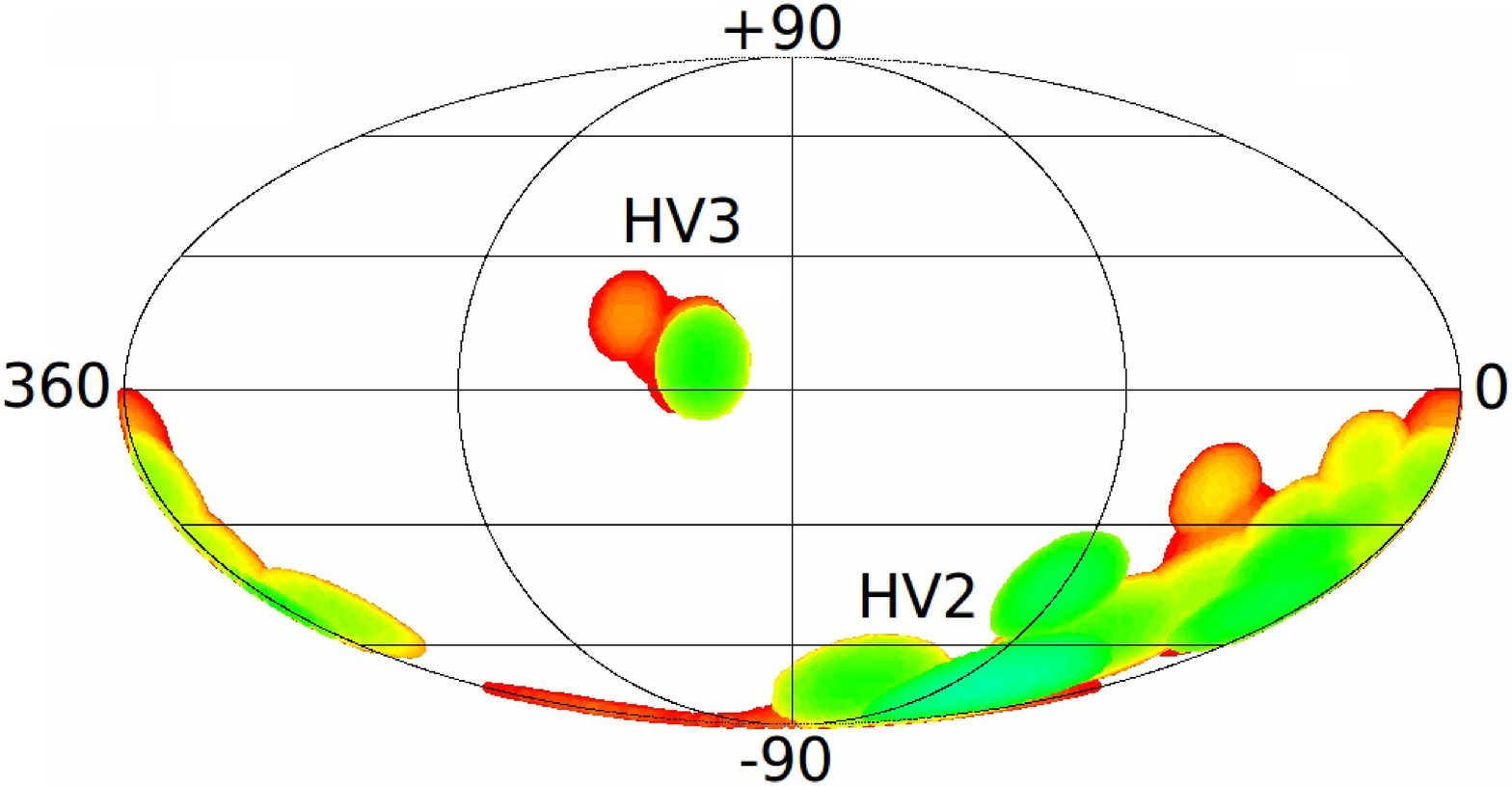}\\
\includegraphics[scale=0.3]{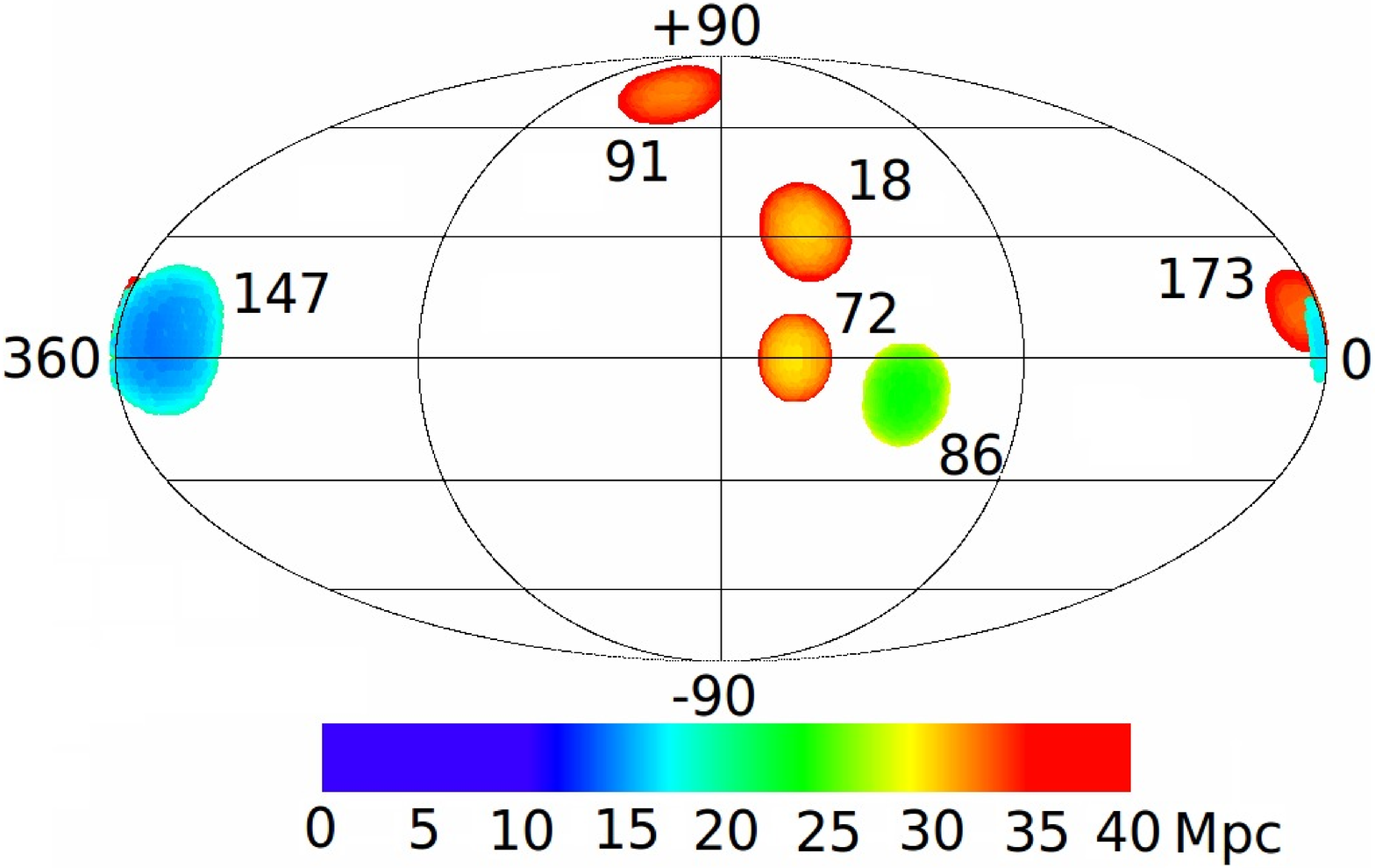}
\caption{Distribution of
voids projected on the sky in the equatorial coordinate system:
the HV1 hypervoid (the top panel), the HV2 and HV3 hypervoids (the
middle panel), individual voids (the bottom panel). The sizes of
the circles correspond to the actual angular scales of voids. Colors show the distances to the surface of voids.}
\end{figure}

As follows from the last column of Table~1, 83  of   89~voids
overlap with each other, forming three groups of hypervoids: HV1,
HV2 and HV3 with 56, 22 and 5 spherical voids, respectively. Some
parameters of these hypervoids: the total volume, the minimum and
maximum distance of the hypervoid surface to the observer, the
distance to the centroid of the hypervoid and its position in the
sky are listed in Table~2. The nearest hypervoid HV1 is actually
an extended and curved bunch of empty  spherical volumes, which
starts just beyond the boundary of the Local Group  \mbox{($D_{\rm
min}=1.4$~Mpc).} The top panel of Fig.~2 represents the   sky
distribution of 56 voids, contained in HV1. The sizes of circles
correspond to the angular scales of the spherical voids. Not to
overshadow the other voids, the contour of the nearest void No.~38
(the front part of the Tully Local Void~\cite{Tul1987:Elyiv_n}) is
described by a thick dashed line.

The HV1 hypervoid, starting from the Local Group in the
\mbox{Hercules--Aquila} region, reaches the boundary of the
considered volume and, passes round the Local Volume in  a
horseshoe shape, and approaches it from \mbox{Gemini--Leo.} The
horseshoe shape of the low-density regions, covering the Virgo
cluster is clearly visible in Fig.~6 of Courtois et
al.~\cite{Cour2012:Elyiv_n}.

Fig.~3  shows the  HV1 in more detail in three projections with
respect to the supergalactic plane. Since the hypervoid has a
complex structure, we show it from the direction of negative and
positive axes perpendicular to the considered plane, in the left
and right panels of the figure, respectively. The distances to the
particular plane are characterized by the scale under the figure.
The upper, middle and bottom panels show the projection on the
SGX--SGY, SGX--SGZ, SGY--SGZ planes, respectively. Concentric
circles have an increment of 10 Mpc. We can see that HV1 departs
from the supergalactic plane quite far away, up to 40 Mpc. In the
SGX--SGZ and SGY--SGZ projections we can clearly see that the
hypervoid envelopes the Local Group. Comparing the projections of
the HV1 hypervoid in Fig.~3 with the corresponding maps of the
Local and Virgo voids from Courtois et
al.~\cite{Cour2012:Elyiv_n}, we can conclude that about 2/3 of the
HV1 volume coincide with the total volume of Local and Virgo
voids.

\begin{table}
\caption{The properties of three local hypervoids}\medskip
\begin{tabular}{l|c|c|c} \hline
&\multicolumn{3}{c}{Hypervoids}\\
\cline{2-4}
        &        HV1     &    HV2    &   HV3   \\
\hline
Number of voids   &        56      &    22     &    5    \\
Volume, Mpc$^3$   &      68469     &   23767   &   3440  \\
 $D_{\rm min}$, Mpc   &        1.4     &    19.4   &   23.8  \\
 $D_{\rm cen}$, Mpc   &       13.8     &    30.5   &   35.8  \\
 $D_{\rm max}$, Mpc   &       46.9     &    47.6   &   45.6  \\
 $\rm RA_{c}$,  deg&       22.2     &     2.4   &   14.2  \\
 $\rm Dec_{c}$,  deg&       $+32$      &   $-51$    &    $+10$  \\
 Sky region       &      Pegasus   &  Eridanus &  Bootes \\
\hline
\end{tabular}
\end{table}

\begin{figure*}[p]
\includegraphics[scale=0.25]{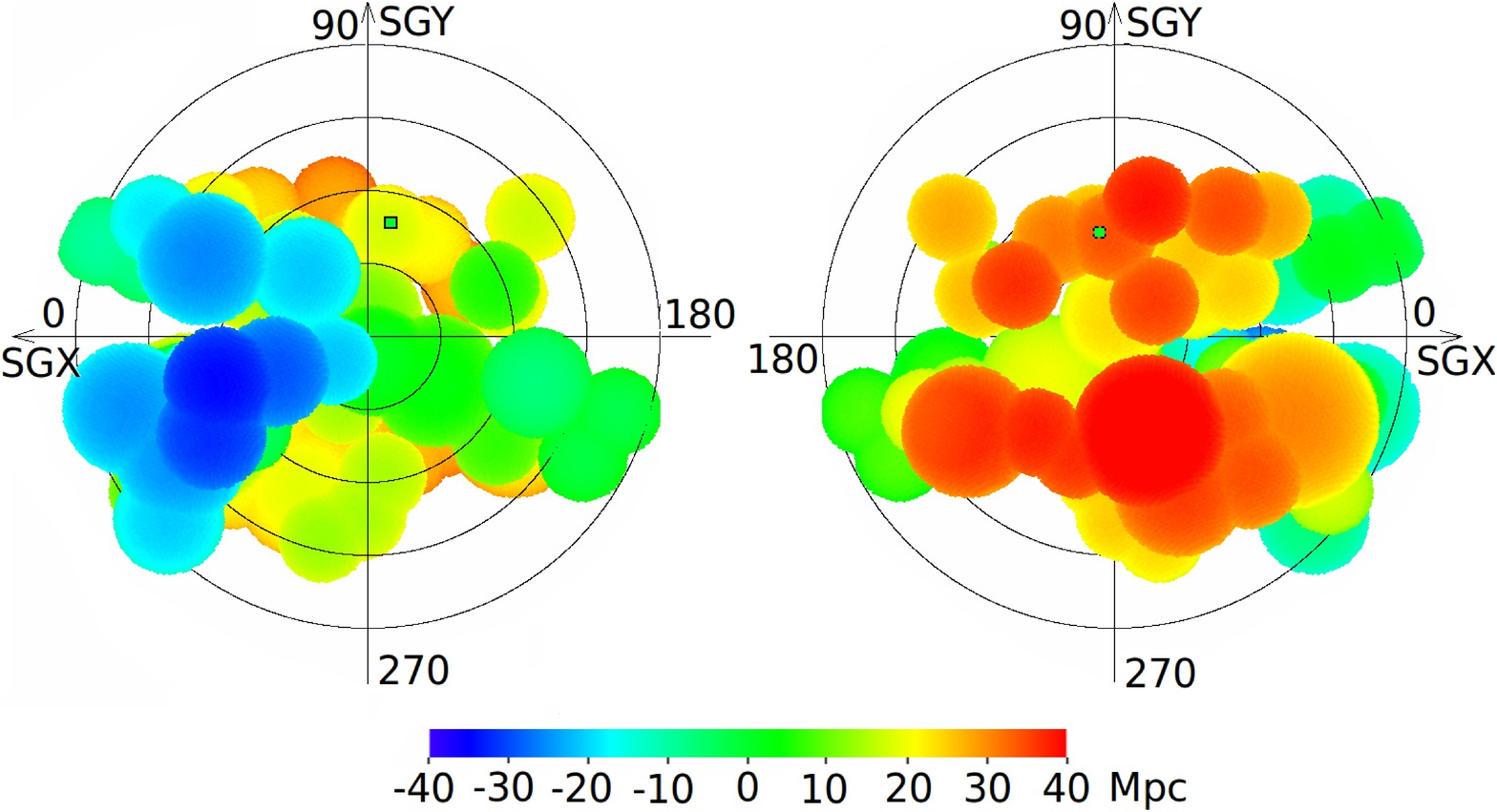}
\includegraphics[scale=0.25]{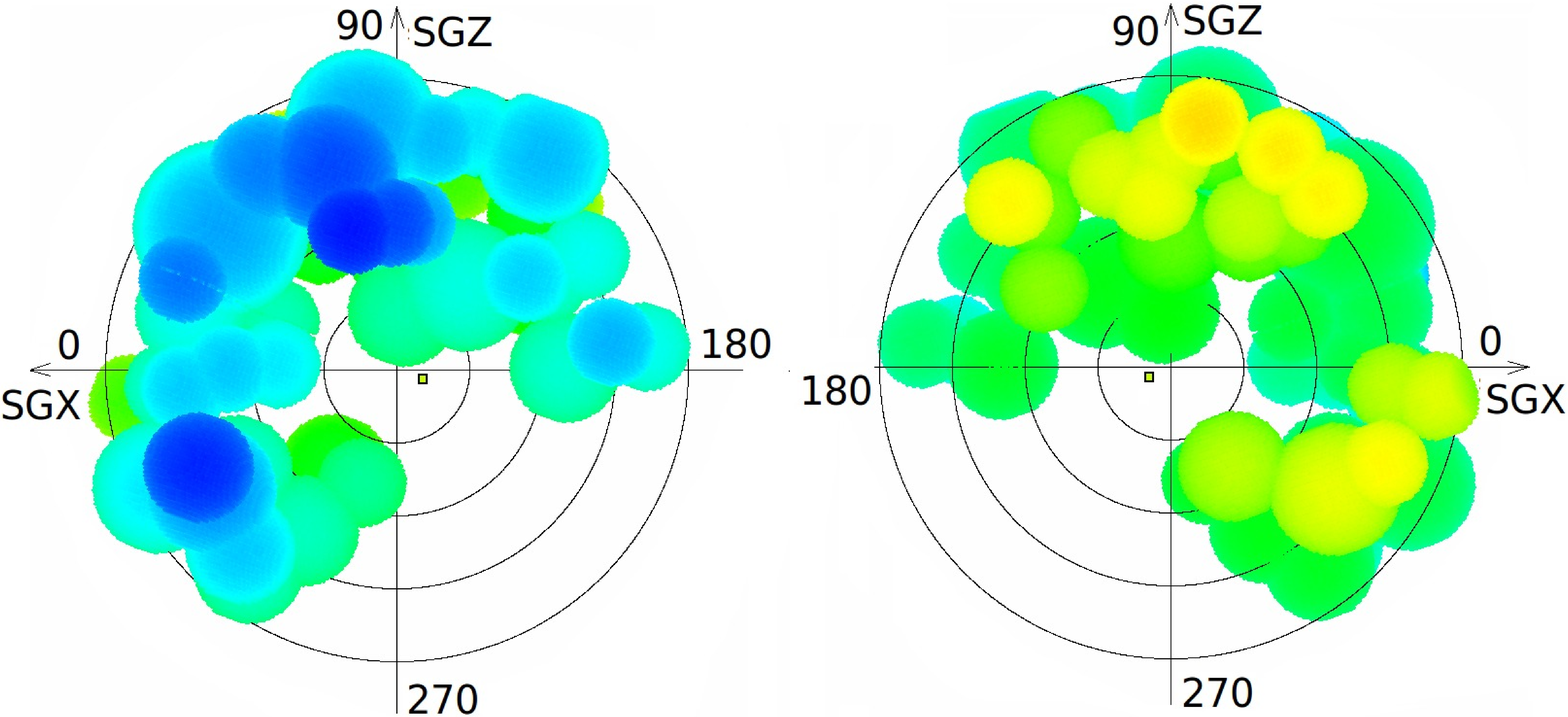}
\includegraphics[scale=0.25]{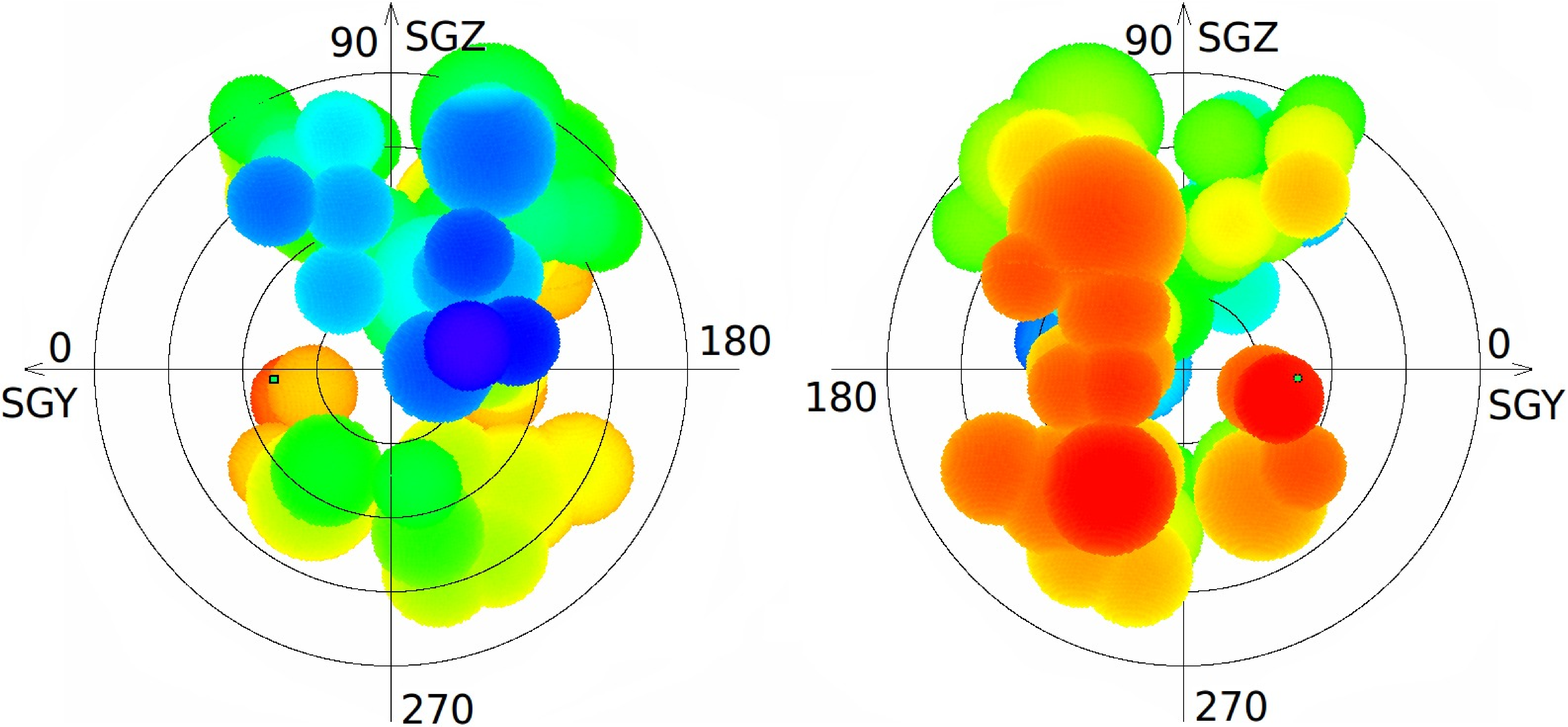}
\caption{Projection of the HV1
hypervoid on the supergalactic SGX--SGY plane (the top panel),
SGX--SGZ (the middle panel), SGY--SGZ (the bottom panel). The view
from the negative (left) and positive (right) directions of the Z,
Y, X axes is shown in three panels, respectively. The square marks
the center of the Virgo cluster. Colors show the distances to the supergalactic plane.} 
\end{figure*}

The middle panel of Fig.~2 reproduces the sky distribution of 22
spherical voids, merging into the HV2 hypervoid in the region of
the Eridanus and 5 voids, belonging to a more compact HV3
hypervoid in the Bootes constellation. The bottom panel of Fig.~2
presents six separate spherical voids, the surfaces of which are
not in contact with other empty volumes, identified by our
algorithm. It is appropriate to recall here that we have
restricted ourselves to the search for voids with linear radii of
at least 6~Mpc. We can assume that there are plenty of small voids
that overlap with the already identified voids, thus increasing
their total volume.

\section{DWARF POPULATION OF NEARBY VOIDS}

Summing up the volume of voids presented in Table~1, we deduce
that they occupy about 30\% of the considered volume of the Local
Universe within 40~Mpc. This estimate takes into account the fact
that the spherical voids belonging to hypervoids overlap.
Searching for nearby voids we have excluded from consideration
2906 dwarf galaxies. If  this population was homogeneously
distributed in the volume of the Local Universe, the expected
number of dwarf galaxies in the voids would amount to about 1000.
Their real number, which is $N=48$ within 40~Mpc and $N=68$ within
48~Mpc does not reach even a tenth of expected. This means that
the empty volumes, devoid of galaxies with normal luminosity
remain almost empty when considering the dwarf galaxies as well.

The list of dwarf galaxies, which are located inside of 89 voids
we have identified, is presented in Table~3. The columns of the
table contain the following information about the galaxies: (1)
the name of the galaxy, or its presence in the SDSS, 6dF, 2MASX,
KUG, HIPASS, APMUKS sky surveys; (2)~the equatorial coordinates
for the epoch J\,2000.0; (3)~the line-of-sight velocity relative
to the centroid of the Local Group (km/s), used to determine the
distance at the parameter $H_0=73$~km/s/Mpc;
(4)~morphological type; (5) apparent $B$-band magnitude; (6)
apparent ultraviolet magnitude $m_{\rm FUV}$, \mbox{($\lambda_{\rm
eff}=1539$~\AA, $\rm{FWHM}=269$~\AA)} according to
GALEX~\cite{Mart2005:Elyiv_n,Gil2007:Elyiv_n}; (7) the flux
\linebreak logarithm $F_{\rm HI}$  in the line of neutral hydrogen
($F_{\rm HI}$ in the \mbox{Jy$\times$km/s} units);  (8) absolute
magnitude; (9) the logarithm of the hydrogen mass
$M_{\rm HI}=2.36\times10^5 D^2 F_{\rm HI}$, where $D$
is the distance in Mpc; (10) the logarithm of the star formation
rate $\log{\rm SFR} = 2.78 - 0.4 m^{\rm c}_{{\rm FUV}}+2\log
D$ in the $M_{\odot}$/yr units, where $m^{\rm c}_{{\rm FUV}}$ is
the UV value, corrected for absorption~\cite{Lee2011:Elyiv_n}, and
the distance $D$ is expressed in Mpc; (11) the depth of bedding of
the galaxy below the surface of the hypervoid (in Mpc); a dash in
this column indicates that the galaxy is on the far edge of the
considered volume  (\mbox{$D>40$}~Mpc), where different edge
effects become significant; (12) notes: the crosses~($\times$)
mark the galaxies from the Catalog of Nearby Isolated Galaxies
LOG~\cite{Kar2011:Elyiv_n}, the pluses~($+$) mark the galaxies
from the list of isolated dwarf galaxies in the Local
supercluster~\cite{Kara2010:Elyiv_n}. The analysis of the data
presented in Table~3 allows us to make the following conclusions.

a)~The distribution of galaxies in the voids based on
morphological types is distinctly shifted towards the latest
types: Im, BCD, Ir, as compared to the samples of dwarfs in groups
and general field. The dwarf galaxies  of the Im, BCD, and Ir
types make up about 65\%, and along with Sm---up to 85\% of the
Table~3 sample. Note that among   the very isolated galaxies of
the LOG catalog~\cite{Kar2011:Elyiv_n} the irregular, BCD, and Sm
objects represent about 51\%. Only in two galaxies,
\mbox{J\,0817+24}~(dEn type) and \mbox{J\,1518--24}~(Sb? type),
their yellowish color indicates the dominance of the old stellar
population. Both of them are located near the void surfaces and
have neighbors with similar line-of-sight velocity values, i.e.
they are members of diffuse groups, adjoining the voids. The
excess of irregular blue galaxies and isolated galaxies among the
samples of galaxies in the voids was also noted
in~\cite{Hoy2005:Elyiv_n,Pati2006:Elyiv_n,Vav2009:Elyiv_n}.

b)~Among   60 dwarf galaxies in the nearby voids, observed with
the GALEX, the \mbox{$\rm FUV$-fluxes} were not detected in only
four of them.  Two of these galaxies were mentioned above as the
group members and two others are projected on the center of a rich
Coma cluster ($=$\,Abell~1656). The remaining objects of our
sample follow a fairly clear correlation between ${\rm SFR}$

{\tiny
\setcaptionwidth{\linewidth}%
\setcaptionmargin{0mm} %
\medskip

\begin{longtable}{l|c|r|l|c|c|r|c|r|c|c|c}
\caption{Dwarf galaxies in the nearby voids} \\
\hline
\multicolumn{1}{c|}{Name}         &        J\,2000.0    & \multicolumn{1}{c|}{$V_{\rm LG}$}  &  \multicolumn{1}{c|}{Type}     &  $B_T$   &   $m_{\rm FUV}$  &   \multicolumn{1}{c|}{$\log F_{\rm HI}$}&   $M_B$    &\multicolumn{1}{c|}{$\log M_{\rm HI}$}& $\log{\rm SFR}$& Depth &Notes\\
 \hline
\multicolumn{1}{c|}{(1)} &(2)& \multicolumn{1}{c|}{(3)}&\multicolumn{1}{c|}{(4)}& (5)&(6)& \multicolumn{1}{c|}{(7)}& (8)& \multicolumn{1}{c|}{(9)}& (10)& (11)& (12)\\
\endfirsthead

\caption{(Contd.) }\\
\hline
\multicolumn{1}{c|}{Name}         &        J\,2000.0    & $V_{\rm LG}$  &  \multicolumn{1}{c|}{Type}     &  \multicolumn{1}{c|}{$B_T$}   &   $m_{\rm FUV}$  &   \multicolumn{1}{c|}{$\log F_{\rm HI}$}&   $M_B$    &\multicolumn{1}{c|}{$\log M_{\rm HI}$}& $\log{\rm SFR}$& Depth &Notes\\
\hline
\multicolumn{1}{c|}{(1)} &(2)& \multicolumn{1}{c|}{(3)}&\multicolumn{1}{c|}{(4)}& (5)&(6)& \multicolumn{1}{c|}{(7)}& (8)& \multicolumn{1}{c|}{(9)}& (10)& (11)& (12)\\
\hline
\endhead
\hline
\endfoot
\endlastfoot
\hline
ESO149--018    &  000714.5--523712& 1744  &  Sdm   &  15.9 &   17.51 &0.74     & --16.04    & 8.87    &--1.43    &0.5 & $\times$  \\
KK261         &  004058.7--261605& 2726  &  Ir    &  17.6 &   18.20   &0.42     & --15.31    & 8.94    &--1.32    &1.1 &   \\
LSBCF682--01   &  005731.9+102148& 2936  &  SBm   &  17.9 &   19.47  &0.40     & --15.40    & 8.98    &--1.58    & -- &    \\
ESO541--005    &  005918.1--203444& 2006  &  Sdm   &  15.8 &   17.39 &0.91     & --16.49    & 9.16    &--1.23    &0.4 &   \\
UGC00655      &  010401.2+415035& 1084  &  Sd    &  14.4 &   16.30    &1.21     & --16.75    & 8.93    &--1.17    &0.6&  $\times$  \\
LSBGF352--021  &  012658.6--350542& 2068  &  BCD   &  17.5 &   19.18 &0.36     & --14.85    & 8.63    &--1.92    &4.3 & $\times$   \\
UGC01038      &  012747.4+431506& 1473  &  Sm    &  17.0   &   18.67 &0.14     & --14.79    & 8.12    &--1.87    &0.6    \\
SDSS          &  013708.1--003354& 3044  &  Sm    &  17.1 &   18.68  &0.00     & --16.12    &$<8.61$    &--1.36    & --     \\
ESO355--005    &  021839.7--363152& 2399  &  Sm    &  17.2 &   18.45 &0.58     & --15.46    & 8.98    &--1.50    &0.5 & $\times$   \\
ESO298--033    &  022128.7--384814& 2142  &  Im    &  16.8 &   18.53 &0.51     & --15.62    & 8.81    &--1.64    &0.2 &$+$   \\
LCRS          &  025224.5--411633& 3470  &  BCD   &  18.1 &   19.33  &$<0.00$   & --15.37     &$<8.72$    &--1.54    & --     \\
6dF           &  051556.2--364418& 1867  &  Ir    &  16.7 &   18.30   &0.14     & --15.51    & 8.33    &--1.59    &0.3    \\
ESO306--010    &  053415.6--391010& 1921  &  Sm    &  16.7 &   17.91 &0.76     & --15.54    & 8.97    &--1.44    &2.4    \\
ESO554--017    &  053557.2--211451& 1385  &  Sd    &  16.7 &   --    &$<0.00$   & --14.83     &$<7.93$    &  --      &0.6    \\
UGC03672      &  070627.6+301919& 964   &  Im    &  15.4 &   17.23   &1.23     & --15.52    & 8.84    &--1.62    &1.6& $\times$$+$    \\
UGC03876      &  072917.5+275358& 811   &  Scd   &  13.7 &   --      &1.06     & --16.72    & 8.52    &  --      &0.3&$\times$    \\
SDSS          &  080158.9+212219& 1343  &  Ir    &  17.7 &   19.69   &$<0.00$   & --13.84     &$<7.90$    &--2.40    &6.1    \\
LCSBS1123P    &  081715.9+245357& 1832  &  dEn   &  17.3 &  $>$23    &0.38     & --14.88    & 8.55    &--3.48    &0.5&    \\
SDSS          &  082712.8+265127& 1779  &  Im    &  17.4 &   --      & --      & --14.76    &  --      &  --      &0.7&$\times$    \\
SDSS          &  083641.1+051625& 2933  &  BCD   &  17.8 &   18.94   &$<0.00$   & --15.34     &$<8.58$    &--1.50    & --&     \\
2MASX         &083735.5+074831& 1280  &  BCD   &  16.7 &   17.93   &0.43     & --14.66    & 8.29    &--1.80    &1.3 & $^*$  \\
CAM0840+1044  &  084236.6+103314& 3437  &  BCD   &  17.6 &   19.21   &$<0.00$   & --15.98     &$<8.72$    &--1.39    & --  &   \\
SDSS          &  091001.7+325660& 1388  &  Ir    &  17.1 &   18.95   & --      & --14.39     &  --      &--2.17    &1.0  &  \\
2MASX         &  091448.8+330115& 1446  &  BCD   &  16.8 &   18.52   & --      & --14.76    &  --      &--1.97    &0.5   & \\
KUG1028+412   &  103118.4+410226& 2568  &  BCD   &  17.6 &   --      & --      & --15.18    &  --      &  --      &0.4 &$\times$   \\
SDSS          &  103950.9+564403& 1216  &  Ir    &  17.5 &   17.87   & --      & --13.64    &  --      &--1.90    & --&     \\
HS1059+3934   &  110209.9+391846& 3267  &  BCD   &  17.7 &   18.51   & --      & --15.63    &  --      &--1.26    & --&     \\
APMUKS        &  110541.0--000602& 3160  &  Sm    &  18.0   &   19.47&$<0.00$   & --15.42     &$<8.64$    &--1.55    & -- & $^{**}$   \\
SDSS          &  112149.2+585434& 1596  &  Ir    &  17.0   &   18.65 & --      & --14.73    &  --      &--1.98    & --   &  \\
SDSS          &  124459.3+525203& 2808  &  BCD   &  18.4 &   19.45   & --      & --14.59    &  --      &--1.78    & --  &   \\
ABELL1656:3237&  125941.3+275015& 3224  &  BCD   &  19.7 &  $>$23    & --      & --13.57    &  --      &--3.10    & --    & \\
ABELL1656:2538&  130040.1+274851& 3495  &  BCD   &  20.3 &  $>$23    & --      & --13.14    &  --      &--3.03    & --   &  \\
SDSS          &  130905.4+134819& 3223  &  Ir    &  18.0   &   20.16 &--0.38   & --15.32    & 8.28    &--1.92    & --   &  \\
SDSS          &  131011.7+135116& 3279  &  BCD   &  18.1 &   20.51   &--0.25   & --15.26    & 8.42    &--2.04    & --  &   \\
SDSS          &  133753.5+635510& 2813  &  BCD   &  17.8 &   20.52   &  --     & --15.21    &  --      &--2.19    &1.0&    \\
SDSS          &  135031.2--013758& 2366  &  Im    &  19.0   &   19.80 &$<0.00$   & --13.80     &$<8.39$    &--1.93    &0.8 &   \\
KKR2    &  140626.9+092133& 3213  &  Sm    &  17.6 &   19.17   &0.26     & --15.74    & 8.92    &--1.51    &0.6  &$+$  \\
SDSS          &  151454.6+341439& 2910  &  Im    &  18.3 &   19.66  & --          & --14.79   &  --      &--1.81    & --  &   \\
2MASX         &  151844.7--241051& 1881  &  Sb?   &  16.6 &  $>$23  & --          & --16.13   &  --      &--3.08    &0.9 &   \\
SDSS          &  151939.3+385255& 3122  &  BCD   &  17.5 &   19.44  & --          & --15.73   &  --      &--1.68    & -- &    \\
SDSS          &  152013.6+400301& 2823  &  BCD   &  17.8 &   19.55  & --          & --15.22   &  --      &--1.80    & -- &    \\
SDSS          &  152644.5+403448& 2890  &  BCD   &  17.8 &   20.16  & --          & --15.26   &  --      &--2.03    & -- &    \\
KKR26         &  161644.6+160509& 2347  &  Im    &  17.8 &   19.03  &0.45         & --14.93   & 8.83    &--1.67    &0.8 &$\times$$+$   \\
LSBCF585--V01  &  162558.6+203949& 2106  &  Ir    &  17.9 &   18.98 &0.11         & --14.65   & 8.40    &--1.70    &4.1& $\times$$+$    \\
SDSS          &  163424.7+245741& 1131  &  BCD   &  18.1 &   20.40   &$<0.00$      & --13.03     &$<7.75$    &--2.86    &0.4&$\times$    \\
SDSS          &  170517.4+355222& 1184  &  Im    &  17.5 &   18.69  & --          & --13.65   &  --      &--2.20    &0.7 &   \\
HIPASS1752--59 &  175251.4--594049& 2596  &  Ir    &  17.2 &   --   &0.52         & --15.89   & 8.99    &  --      &0.6 &   \\
UGC11109      &  180414.0+464414& 1820  &  Sm    &  17.2 &   18.25  &0.72         & --14.97   & 8.88    &--1.58    &3.9 &$\times$   \\
UGC11220      &  182325.5+405643& 1706  &  Sm    &  16.7 &   17.00     &0.98         & --15.34   & 9.08    &--1.13    &1.7 &$\times$   \\
HIPASS1926--74 &  192727.1--740458& 2444  &  BCD   &  17.0   &   -- &0.36         & --15.91   & 8.78    &  --      &2.4  &$\times$  \\
KK246         &  200357.4--314054& 572   &  Ir    &  17.1 &   20.01 &0.90         & --13.70   & 8.06    &--2.43    &4.5 &$\times$$\lefteqn{^{***}}$   \\
6dF           &  210804.9--471941& 832   &  BCD   &  15.9 &   18.12 &$<0.00$      & --14.53     &$<7.48$    &--2.24    &0.8 &   \\
LSBCF743--01   &  211845.4+082202& 3203  &  Sm    &  17.5 &   18.13 &0.46         & --16.00   & 9.11    &--0.96    &0.6  &  \\
CGCG426--040   &  212006.0+115506& 1415  &  BCD   &  16.4 &   18.64 &0.34         & --15.41   & 8.28    &--1.82    &1.7 &   \\
SDSS          &  212202.3+095311& 3237  &  BCD   &  17.7 &   --     &$<0.00$      & --15.77      &$<8.66$    &  --      &0.4 &   \\
ESO531--001    &  213152.0--235632& 2668  &  Sm    &  17.1 &   18.47&0.09         & --15.92   & 8.59    &--1.32    & --  &   \\
UGC11771      &  213527.5+232805& 1951  &  Sd    &  16.4 &   18.60   &0.64         & --16.17   & 8.86    &--1.47    &0.5&    \\
UGC11813      &  214731.1+220951& 2124  &  Sm    &  17.3 &   19.07  &0.71         & --15.51   & 9.01    &--1.54    &1.5&    \\
SDSS          &  223036.8--000637& 1758  &  BCD   &  17.4 &   --    &--0.22       & --14.81   & 7.92    &  --      &0.4&    \\
ADBS          &  225558.3+261011& 2930  &  BCD   &  17.7 &   19.21  &0.35         & --15.79   & 8.93    &--1.33    &0.9&    \\
LSBCF469--02   &  225721.5+275852& 3233  &  Sm    &  18.3 &   18.25 &0.46         & --15.21   & 9.12    &--1.01    & -- &$\times$    \\
SDSS          &  230511.2+140346& 1801  &  BCD   &  17.3 &   19.38  &$<0.00$      & --15.52     &$<8.15$    &--1.52    &0.3 &   \\
6dF           &  231803.9--485936& 2275  &  BCD   &  16.9 &   18.42 &$<0.00$      & --15.62     &$<8.36$    &--1.56    &3.8  &$\times$  \\
KKR75         &  232011.2+103723& 1703  &  Ir    &  18.0 &   19.48  &0.54         & --14.05   & 8.65    &--2.11    &1.9 &$\times$$+$   \\
LSBCF750--04   &  234420.2+100705& 1726  &  Sd    &  17.3 &   18.62 &0.40         & --14.80   & 8.51    &--1.74    &0.7 &   \\
UGC12771     &  234532.7+171512& 1535  &  Im    &  16.9 &   18.35  &0.46         & --14.96   & 8.48    &--1.72    &0.4 &$\times$$+$   \\
APMUKS        &  234650.9--301106& 2926  &  BCD   &  18.2 &   19.61 &$<0.00$      & --14.89     &$<8.58$    &--1.80    &0.7    \\
LSBCF750--V01  &  235419.6+105636& 1164  &  Ir    &  18.0   &   19.66&$<0.00$     & --13.41     &$<7.78$    &--2.37    &0.4    \\
\hline 
\end{longtable}}

{\scriptsize
\noindent Notes:\\
$*$ The LEDA specifies a significant difference in the
heliocentric velocity estimates  of this galaxy from the SDSS
optical data    ($+1452\pm19$ km/s) and from the HI measurements
in the HIPASS   ($+2006\pm8$ km/s). Re-processing of the optical
spectrum yields $+2004\pm15$ km/s, which is close to the HIPASS
estimate.   \\
$**$ The line-of-sight velocity value of  this the galaxy,
 obtained in~\cite{Imp2001:Elyiv_n}  needs to
be confirmed.\\
$***$ Column (3) indicates the formal value of the line-of-sight
velocity, corresponding to the distance of the galaxy
7.83~Mpc~\cite{Kar2006:Elyiv_n} and the parameter $H_0=73$
km/s/Mpc. Having the line-of-sight velocity of $V_{\rm LG}=+436$
km/s, this galaxy is moving to us from the depth of the void with
the  peculiar velocity of --130~km/s.}\\

\begin{figure}[b]
\vspace{-8mm}
\includegraphics[width=0.96\columnwidth]{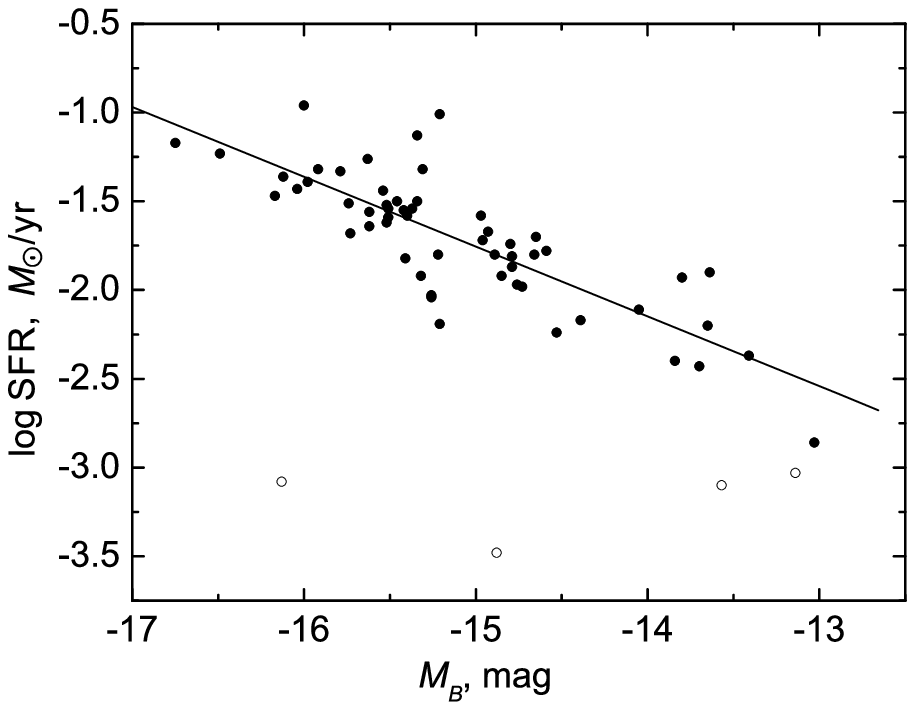} \vspace{-7mm}
\caption{Relationship between the
logarithm of star formation rate ${\rm SFR}$ and absolute
magnitude  $M_B$ for  60  dwarf galaxies in the voids. The
unfilled circles show the galaxies with upper estimates of
$m_{\rm FUV}$, which were not accounted for in the regression
parameters: \mbox{$\log{\rm SFR}=-0.39 M_B-7.65$},
\mbox{$R=-0.82$}, \mbox{${\rm SD}=0.22$}, where $R$ is the
correlation coefficient, and ${\rm SD}$ is the RMS deviation.}
\end{figure}

\begin{figure}[b]
\vspace{-8mm}
\includegraphics[width=0.96\columnwidth]{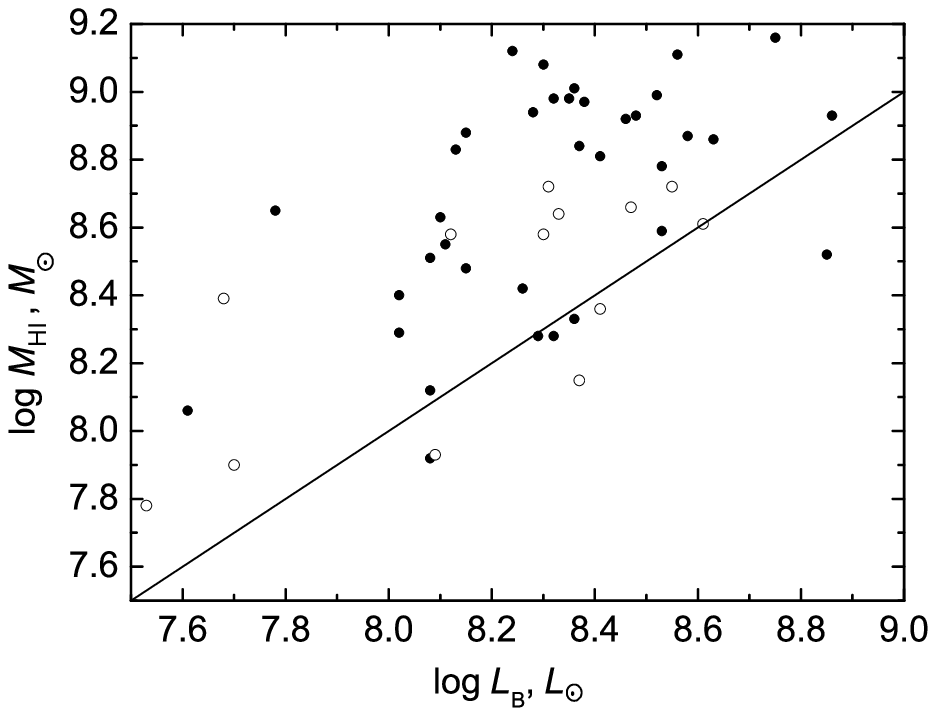} \vspace{-7mm}
\caption{Distribution of 51 dwarf
galaxies in the voids by hydrogen mass logarithm and
$B$-luminosity. Open symbols mark the  objects with an upper limit
estimate of  $M_{\rm HI}$. The straight line corresponds to
\mbox{$M_{\rm HI}/L_B = 1~M_{\odot}/L_{\odot}$.}} \vspace{9.5mm}
\end{figure}

\noindent  and absolute magnitude of the galaxy (see Fig.~4) with
the \mbox{$R=-0.82$} correlation coefficient and the median value
of the specific star formation rate \mbox{${\rm SSFR} = 1\times
10^{-10}~M_{\odot}$\,yr$^{-1}L_{\odot}^{-1}$,} typical of the
late-type dwarfs in the Local Volume~\cite{Kar2012:Elyiv_n}.

c)~The distribution of dwarf galaxies in the voids based on the
hydrogen mass logarithm and \mbox{$B$-luminosity} is
shown in Fig.~5 by the solid circles. The upper limits of the
hydrogen mass estimates $M_{HI}$ are shown by unfilled circles. As
follows from these data, the dwarfs in the voids have a high
hydrogen abundance per luminosity unit. The median value for them,
\mbox{$M_{\rm HI}/L_B=2.1~M_{\odot}/L_{\odot}$} proves to be
approximately three times higher than for the Ir, Im, and
\mbox{BCD} galaxies in the groups and general
field~\cite{Kar2004:Elyiv_n}. In other words,  dwarf galaxies in
the voids possess increased gas reserves compared to the galaxies
of the same type located in denser environments, what was
repeatedly mentioned by different
authors~\mbox{\cite{Pus2011:Elyiv_n,Kara2010:Elyiv_n,Huch1997:Elyiv_n,Pust2002:Elyiv_n,Pust2011:Elyiv_n,Stan2010:Elyiv_n,Krecl2012:Elyiv_n}.}
Possessing normal  star formation rates (SFR) per luminosity unit,
$\log{\rm SSFR}\simeq-10$,  the dwarfs in the voids are able to
maintain the observed SFR on the scale of about 20--30~Gyr. At the
same time, the studies~\cite{Pati2006:Elyiv_n,Sor2006:Elyiv_n}
considering the properties of galaxies in the voids in much larger
\mbox{volumes~($z<0.09$)} demonstrate that although in general the
void galaxies are bluer and fainter than the cluster galaxies,
however, in the same luminosity range  no differences in colors or
SFRs are observed.

d)~Figure 6  shows the distribution of 48 galaxies in the nearby
($D<40$ Mpc) voids by the absolute magnitude and the depth of
bedding under the hypervoid's surface. We can see that only a
quarter of these galaxies  are located in the voids at  depths
exceeding 1.5~Mpc. Recall that in our algorithm the accuracy with
which the  position of the  spherical void center was determined
was exactly 1.5~Mpc. Consequently, plenty of dwarf galaxies in
this boundary layer can be located outside the voids. In any case,
the population of dwarf galaxies in the cores of voids (${\rm Depth} > 1.5$~Mpc) is represented literally by single
objects, such as their nearest and most famous representative
KK\,246~\cite{Beg2008:Elyiv_n,Kreck2011:Elyiv_n,Gent2012:Elyiv_n}.
Note that only 4 out of 13 galaxies that are located in the middle
of voids are ``new''. The nine remaining objects are specified as
isolated in the lists of~\cite{Kar2011:Elyiv_n,Kara2010:Elyiv_n}.
Since the selection criteria in this and the  two above studies
were completely different, we have to expect that these nine
galaxies are very isolated objects. In Fig.~6 and Table~3 all the
galaxies, identified with the objects from the lists
of~\cite{Kar2011:Elyiv_n,Kara2010:Elyiv_n} are marked by the corresponding signs.

\begin{figure}[]
\includegraphics[width=0.96\columnwidth]{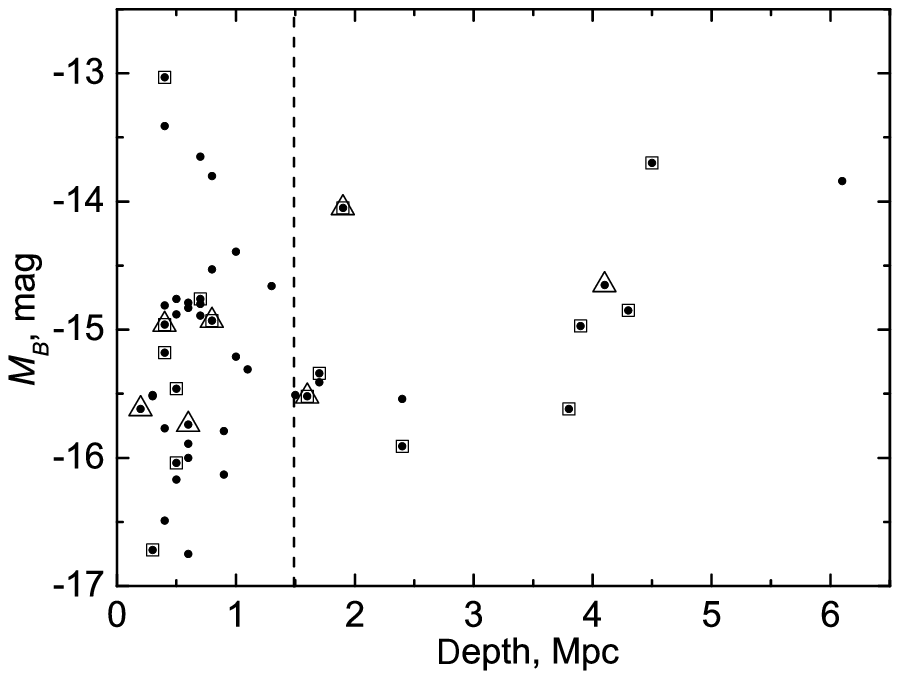} \vspace{-6mm}
\caption{Distribution of 48 dwarf
galaxies in the nearby ($D<40$ Mpc) voids by absolute magnitude
$M_B$ and the bedding depth under the surface of hypervoid.
Squares trace the objects common with the galaxies from the
Catalog of Nearby Isolated Galaxies~\cite{Kar2011:Elyiv_n}, 
triangles denote the  late-type dwarf galaxies from the list
of~\cite{Kara2010:Elyiv_n}.}
\end{figure}

Despite the poor statistics, note a certain tendency of declining
dwarf galaxy luminosity with increasing depth below the hypervoid
surface. The same feature was noted earlier  by Chengalur and
Pustilnik~\cite{Chen2012:Elyiv_n}. From general considerations we
may assume that such objects have record low metallicities. This
makes them interesting for the spectral observations.

e)~All the dwarf galaxies we have discovered in the nearby voids
have absolute magnitudes brighter than $-13^{\rm m}$. Their
distribution by distance and absolute magnitude in Fig.~7 shows
that this limit may be due to the distance selection effect.
However, the nature of the data in Fig.~7 does not contradict the
assumption about the existence of a luminosity threshold,
$M_B\simeq-13.0^{m}$ in the dwarf population of voids. This
circumstance may have a vital importance for understanding the
nature of empty cosmic volumes.

\section{CONCLUDING REMARKS}

To search for empty volumes in the Local Universe, we used an
algorithm similar to the approach of Patiri et
al.~\cite{Pat2006:Elyiv_n}, only with a much more stringent
restriction on the luminosity of galaxies, evading the voids. As a
result we have obtained a list of nearby spherical voids with the
typical diameter of 15~Mpc, which is about three times smaller
than that in~\cite{Pat2006:Elyiv_n}. However, the total volume of
our voids is about 30\% of the volume of the Local Universe within
40~Mpc. The distribution of the centers of spherical voids proved
to be very far from the Poisson distribution. More than 90\% of
the voids overlap, forming three hypervoids composed of
56,~22~and~5 initial voids, respectively. The closest and most
populated hypervoid HV1 comprises the Tully Local Void and extends
in a horseshoe shape, covering the central region of the Local
Supercluster of galaxies.

\begin{figure}[]
\includegraphics[width=0.96\columnwidth]{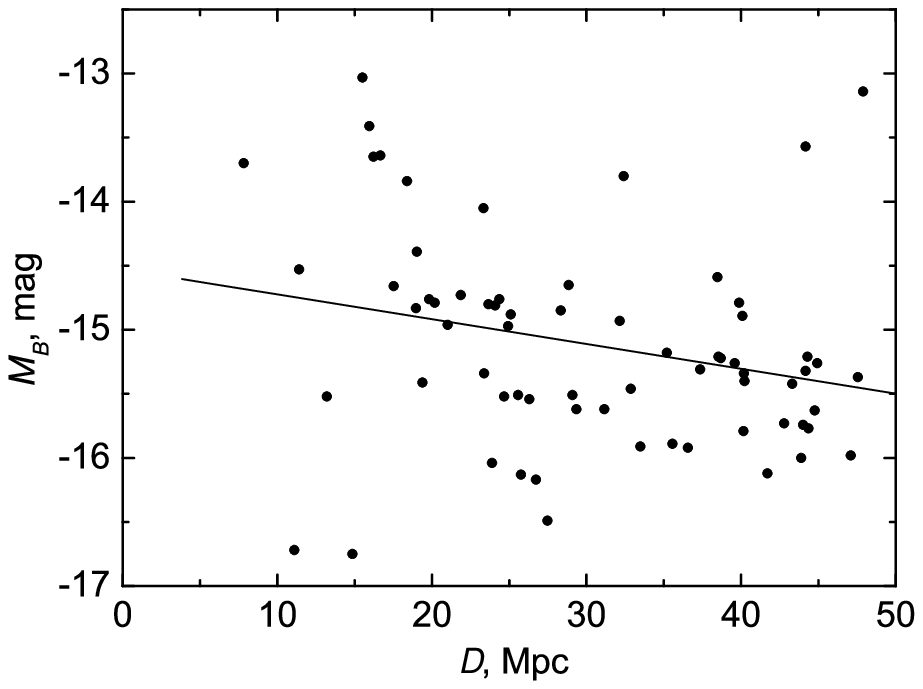}  \vspace{-6mm}
\caption{Distribution of 68 dwarf
galaxies by absolute magnitude $M_B$ and distance from the
observer. The regression parameters are as follows:
\mbox{$M_B=-0.02 D-14.53, R=-0.26, SD=0.8$}.}
\end{figure}

Eighty-nine voids we have identified with the center distances
within $<40$~Mpc from us contain 48 late-type dwarf galaxies with
absolute magnitudes in the range of $M_B=[-13.0, -16.7]$. These
galaxies have active star formations and gas reserves per
luminosity unit by about 2--3 times higher than those in the dwarf
galaxies of the same type, located in denser environments.  The
abundance of neutral hydrogen in void galaxies was repeatedly
noted by different
authors~\cite{Pus2011:Elyiv_n,Kara2010:Elyiv_n,Huch1997:Elyiv_n,Pust2002:Elyiv_n,Pust2011:Elyiv_n,Stan2010:Elyiv_n,Krecl2012:Elyiv_n}.
The void dwarfs reveal a tendency of evading the depths of the
voids and hypervoids. In fact, the central regions of voids are
devoid of not only normal, but also dwarf galaxies. According to a
rough estimate, in the hearts of the voids having the sizes of
about a half of the diameter, the average stellar mass density is
two orders lower than the average cosmological density.

We should note here the following important fact. We have defined
the contours of the nearby voids and identified the dwarf
population there in the space of radial velocities rather than
that of actual distances. The presence of collective motions of
galaxies with the amplitudes of 300~km/s can lead to a significant
distortion of void shapes and the global pattern of their
distribution. Obviously, this situation will gradually become
clearer as more and more data is obtained on the individual 
distances of galaxies.

As we can see from Table~3, about one-third of dwarf galaxies in
the nearby voids ensue from  the photometric and spectral sky
survey SDSS~\cite{Abaz2009:Elyiv_n}. An extension of this fruitful
survey to the other remaining regions of the northern sky, as well
as a prospective similar mass survey in the southern sky will soon
allow us to better understand the characteristics of the dwarf
population of voids, in particular, to find out whether there
exist  any ultra-faint dwarf objects with luminosities fainter
than $3\times10^7~L_{\odot}$ and hydrogen masses lower than
$10^7~M_{\odot}$ in the empty cosmic volumes.\\
\\

{\bf Acknowledgments}\\
The present study has made use of the SDSS ({\tt
http://www.sdss.org}), HyperLeda \\
 ({\tt
http://leda.univ-lyon1.fr}) and NED ({\tt
http://nedwww.ipac.caltech.edu}) databases. This study was made
owing to the support of the following grants: the grants of the
Russian Foundation for Basic Research (project nos.
\mbox{12-02-91338-NNIO}, \mbox{11-02-00639},
\mbox{11-02-90449-Ukr-f-f}, the State Foundation for Basic
Research of the Ukraine F40.2/49, the Cosmomicrophysics program of
the National Academy of Sciences of the Ukraine, as well as by the
Ministry Education and Science of the Russian Federation (state
contract no.~14.740.11.0901) and project
\mbox{2012-1.5-12-000-1011-004}.

{}

\end{document}